\documentstyle[11pt,jd24]{article}

\begin{document}

\title{Infrared Observations of Large Amplitude Pulsating Stars }
\author{Patricia A. Whitelock}
\affil{South African Astronomical Observatory, P O Box 9, Observatory,
7935 South Africa. e-mail: paw@saao.ac.za}

\begin{abstract} Our understanding of large amplitude pulsating stars and
their status in stellar evolution is briefly reviewed. The paper then
describes the near-infrared light curves of various asymptotic giant branch
stars, concentrating on possible evidence for changing mass-loss rates. The
stars discussed include oxygen- and carbon-rich Miras, OH/IR stars,
thick-shelled carbon stars and symbiotic Miras. Finally a newly discovered
Mira variable in the Sagittarius Dwarf Galaxy is described. \end{abstract}


\section{Introduction}
 The following discussion concentrates on asymptotic giant branch (AGB)
stars, including both classical Miras and the longer period infrared sources
which seem to be closely related to the Miras. Although many supergiants
also fall in the large amplitude variable category they are not considered
here. The Miras have large amplitudes by definition; their visual magnitude
range exceeds 2.5 mag, although IR and bolometric amplitudes are usually
smaller. They are long period, mostly in excess of 100 day and have spectra
dominated by molecular absorption with hydrogen lines in emission at certain
phases (Kholopov et al.\ 1985, GCVS). The infrared variables overlap in
properties with the Miras, but their periods extend to over 2000 days. They
have very late spectral types and although they are often too faint to have
measured visual amplitudes, near-IR amplitudes of a magnitude or more
are common (e.g.\ Whitelock, Feast \& Catchpole 1991).

\section{Stellar Evolution}
 We understand these stars to be near the end of their AGB evolution (Iben
\& Renzini 1983). Their next evolutionary move will be across the HR diagram
to become white dwarfs, possibly via a planetary nebula phase. Chemical
enrichment, in particular the presence of technetium (Merrill 1952; Little
et al.\ 1987) tells us that many of them are on the thermally pulsing AGB
and experiencing third dredge-up. From the strong thermal infrared flux we
deduce that they are losing mass copiously and are probably in the {\it
super-wind} evolutionary phase identified by Renzini (1981). They obey a
period-luminosity (PL) relation (e.g.\ Feast et al.\ 1989; Whitelock et al.\
1994) making them potentially useful distance indicators. The longer period,
higher luminosity stars presumably had a higher initial mass than their
short period counterparts although there is probably not a one-to-one
relationship of period to initial mass across the whole period range. At the
short period end, 200 day Miras are found in metal-rich globular clusters
(Menzies \& Whitelock 1985) so we know more about them than the others. As
expected, the kinematics of the short period Miras differ from those of the
longer period stars (Feast 1963; Whitelock et al.\ 1994)

\section{Pulsation and Mass loss}
 A number of interesting theoretical papers have appeared recently
concerning the connections between pulsation and mass loss. Of particular
note are those by the Vienna group (e.g.\ H\"ofner, Feuchtinger \& Dorfi
1995) and by the Berlin group (e.g.\ Fleischer, Gauger \& Sedlmayr 1995),
both of whom describe and model the dust-induced $\kappa$-mechanism which
leads to the periodic formation of dust and creation of shock waves in
material flowing outwards from a mass-losing star. Thus a theoretical
framework for the interpretation of accurate light curves has become
available for the first time. One paper in particular, Winters et al.
(1994), makes detailed predictions of the IR light curves of mass-losing
carbon stars.  Their models predict the periodic formation of dust on a
timescale which is generally some multiple of the stellar pulsation period. 
Thus their predicted light curves show the results of the superposition of
two oscillations, that of the interior pulsation and that produced by the
dynamical structure of the circumstellar dust shell. More recently,
H\"ofner \& Dorfi (1997) have pointed out that the behaviour of the dust is
not periodic for most of their models. This seems important as it is
more consistent with the observations which are described below.

\begin{table*}[b]
\centering
\caption{Variable Stars}
\begin{tabular}{@{}lccccc}
\multicolumn{1}{c}{Star} & \multicolumn{1}{c}{Sp} & 
\multicolumn{1}{c}{Period} & \multicolumn{1}{c}{$\Delta K$} & 
\multicolumn{1}{c}{$\dot{M}$}& symbol\\
& \multicolumn{1}{c}{type} &
\multicolumn{1}{c}{(day)} & \multicolumn{1}{c}{(mag)} & 
\multicolumn{1}{c}{($\times 10^{7}M_{\odot}yr^{-1}$)}\\ 
\hline
S Ind & M & 400 & 0.8 & 2 & $\triangle$\\
R Cae & M & 398 & 0.8 & 2& $\triangle$\\
\hline
IK Tau & M & 470 & 1.0 & 50 & $\circ$\\
$\rm OH327-0.6$ & M & 600 & 1.2 & 60 & $\circ$\\
\hline
R For & C & 389 & 0.9 & 8 & $\bullet$\\
IZ Peg & C & 486 & 2.0 & 200 & $\bullet$\\
\hline
R Aqr & Sym & 387 & 0.8 & 6 & $\times$\\
RR Tel & Sym & 387 & 0.9 & 30 & $\times$\\
\end{tabular}
\end{table*}

\section{Observations}
  The IR (broad-band $JHKL$) photometry discussed below was obtained with
the MkII IR photometer on the 0.75-m telescope at SAAO, Sutherland. The
results are part of a long-term monitoring programme started by Michael
Feast over 20 years ago. All measurements are on the SAAO system as defined
by Carter (1990).

It is particularly useful to work in the 1 to 4 $\mu$m region when dealing
with AGB variables, first, because they are cool stars whose stellar energy
distributions peak in the IR, so that we are examining the behaviour of the
bolometric output rather than minor fluctuations in the temperature;
secondly, the $J$ and $L$ light curves respond rather differently to the
effects of increased dust in the stellar atmosphere. The stars whose light
curves are illustrated below have IR colours which suggest the presence of
circumstellar shells and hence moderate to high mass-loss rates.

The stars under discussion are listed in Table 1 and their mean colours are
illustrated in a two-colour diagram (Fig 1).  Their spectral types (Sp),
pulsation periods (P), $K$ amplitudes ($\Delta K$) and approximate mean
mass-loss rates ($\dot{M}$) are also tabulated. The sample comprises: two
oxygen-rich Miras with periods around 400 day, S Ind and R Cae; two much
redder longer period OH/IR Miras, IK Tau and $\rm OH327-0.6$; two carbon
stars, the bright Mira R For and a dust-shell source, IZ Peg, with a
somewhat longer period and much redder colours; and finally there are two
oxygen-rich symbiotic Miras, R Aqr and RR Tel, with periods around 400 day.
The two oxygen-rich Miras have the bluest colours and therefore the thinnest
shells. All of the other colours show the clear signature of reddening,
presumed to be circumstellar. Only $\rm OH327-0.6$ is in the Galactic Plane
and thus may experience some interstellar reddening, but its bright
apparent luminosity suggests it is close, so even there the major cause of
the red colours must be circumstellar extinction.

\begin{figure}
\plotfiddle{fig1.ps}{8cm}{0}{100}{100}{-200}{-70}
\caption{The IR colours of the stars listed in Table 1 and discussed below.
The symbols are defined in the table.}
\end{figure}

\section{Light Curves}
\subsection{O-rich Miras and OH/IR Stars}

Figure 2 illustrates the $J$ and $L$ light curves for the two oxygen-rich
Miras, $J$ and $L$ being the two extreme wavelengths on the SAAO monitoring
programme.  The data shown here span over 22 yr. Regular pulsations are
clearly evident at both wavelengths, with those at the shorter wavelength
having the larger amplitude, as is normal for most pulsating stars. You can
also see in both figures the same kind of erratic irregularity that
characterises the visual light curves of Miras (e.g.\ the AAVSO curves shown
in Whitelock 1996); cycles do not repeat exactly from one to the next
and there are trends on a variety of longish time scales. Notice in
particular that when the $J$ curve fades so does the $L$ curve; although the
amplitude is smaller at $L$ the behaviour is qualitatively identical. S~Ind
and R Cae also have remarkably similar light curves and specifically similar
long-term variations.

\begin{figure}
\plotfiddle{fig2.ps}{18cm}{0}{110}{110}{-248}{-70}
\caption{The $J$ and $L$ light curves for R Cae and S Ind from 1975 to 1997.}
\end{figure}

Still with the oxygen-rich variables, Fig 3 illustrates the light curves of
the longer period stars IK Tau (NML Tau) and $\rm OH327.4-0.6$, which are
both bright OH Masers (Wilson \& Barrett 1972; Caswell \& Haynes 1975).
Larger amplitude variations are evident at $J$ than at $L$. The $J$ curves
show clear long-term trends in opposite directions for the two stars; IK Tau
is fading and $\rm OH327.4-0.6$ brightening. Although in both cases $L$
follows the same long-term trend as $J$, it does so at a much lower
amplitude. This is in contrast to the previous two stars. These trends are
probably produced by changing dust obscuration, although it is possible that
changes in the stellar temperature could produce the effects. A comparison
of the long-term IR behaviour with that at other wavelengths would be
instructive as would a detailed comparison with model predictions.

\begin{figure}
\plotfiddle{fig3.ps}{18cm}{0}{110}{110}{-248}{-70}
\caption{The $J$ and $L$ light curves for IK Tau and $\rm OH327.4-0.6$ 
from 1979 to 1997.}
\end{figure}

\subsection{Carbon Variables}
 IZ Peg (CRL 3099) was too faint at $J$ near minimum light for reliable
photometry with the 0.75-m telescope, therefore $H$-band photometry is
illustrated in Fig 4. Its extreme colours are the consequence of a thick
circumstellar shell and strong circumstellar reddening. Its light curve is
very similar to that of $\rm OH327.4-0.6$, being fairly regular with some
gradual long-term trends. R For on the other hand, also illustrated in Fig
4, shows rather more dramatic changes in its $J$ light curve. It is in this
respect typical of other carbon stars, with moderate dust shells, e.g.\ R Lep
and R Vol, which have been monitored from SAAO.  These stars show changes of
$\Delta J \sim 2$ mag on time-scales of a few years. The analysis of these
data is discussed elsewhere (Whitelock et al.\ 1997; Whitelock 1997) and will
therefore be described only briefly here.

\begin{figure} 
\plotfiddle{fig4.ps}{18cm}{0}{110}{110}{-248}{-70}
\caption{The $J$ and $L$ light curves for R For from 1975 to 1997 
and the $H$ and $L$ curves for IZ Peg from 1979 to 1997.} 
\end{figure}

The $J$ light curve for R For, after the normal Mira pulsations have been
removed, is illustrated in fig 3 of Whitelock (1997).  What we see in this
residual curve are changes in obscuration of the star, apparently as a
consequence of the erratic production of dust.  Whitelock et al.\ showed that
it is possible to model the brightening event around JD 2446000 as the
consequence of an expanding cloud of dust formed $500\, R_{\odot}$ from the
star and moving with a velocity of $\rm 20\,km\,s^{-1}$. This dust ejection
cannot have been spherically symmetric or the fading at $J$ would have been
accompanied by brightening at $L$, which it was not.  The changes between JD
2447000 and 2450000 could be explained as the effects of a number of
overlapping dust ejections. The dust might thus be ejected as puffs in
random directions as it is in RCB stars (Feast 1997). In which case the star
will ultimately acquires a patchy, but essentially spherically symmetric,
shell. Alternatively, the dust might be ejected preferentially into a torus
or disk which, to explain the R For observations, would have to be in our
line of sight. The torus model would help explain the rather low apparent
luminosities measured for stars like R Lep (van Leeuwen et al.\ 1997;
Whitelock et al.\ 1997).

\subsection{Symbiotic Stars} 
 Particularly good light curve coverage has been obtained for several
symbiotic Miras. These binary stars, comprising a Mira interacting with a
white dwarf, offer us a somewhat different perspective on the mass-loss
problem and we should recall that there must be many unidentified binary
systems among the catalogued Miras. The symbiotic Miras are also known as
{\it D-types}, D for dust, because their IR colours are clearly modified by
the presence of dust (Webster \& Allen 1975; Whitelock 1987) obscuring the
Mira. Their optical and ultraviolet spectra are dominated by high excitation
emission lines, such as HeII and [OIII]. These lines are excited in a
circum-binary nebula formed during mass transfer between the two stars. 
Thus the presence of the Mira only makes itself felt at long wavelengths and
the existence of a cool star in symbiotic systems only became clear
following IR observations. The orbital periods of these stars are unclear
but generally thought to be of the order of decades (Whitelock 1987).

\begin{figure} 
\plotfiddle{fig5.ps}{18cm}{0}{110}{110}{-248}{-70}
\caption{The $J$ and $L$ light curves for the symbiotic Miras R Aqr and
RR Tel from 1975 to 1997.} 
\end{figure}

R Aqr, which is by far the best studied symbiotic Mira, is a bright star
associated with an obvious and extensive reflection nebulosity.  It is also
associated with what is sometimes described as a jet, which is bright both
at ultraviolet and at radio wavelengths. The Mira has a pulsation period of
387 day, and is unusual among symbiotic Miras in that it, and not the hot
component, dominates the optical light. The Mira PL relation, the Hipparcos
parallax and geometrical arguments all imply a distance around 200 to 250 pc
(Whitelock 1987; van Leeuwen et al.\ 1997; Hollis, Pedelty \& Lyon 1997).
Willson et al.\ (1981) suggested a binary period of about 44 yr on the
basis of dips seen in the AAVSO light curve, interpreted as eclipses (see
also Whitelock et al.\ 1983). The ``eclipse'' would be caused by an orbiting
dust cloud obscuring the Mira. One of these ``eclipses'' corresponds to the
fading seen in Fig 5 around JD 2443000. The depth of the ``eclipse'' exceeds
two mag at $J$, but reaches only about one mag at $V$. The later part of Fig
5 looks like a normal Mira light curve. There are the usual apparently
erratic differences from one light cycle to another, but nothing as dramatic
as the ``eclipse''.

The second part of Fig 5 shows another symbiotic Mira, RR Tel, which is also
a very slow nova. It first gained notoriety with its outburst in 1944, when
it apparently brightened by 7 or 8 mag in the visual. It was initially
thought that the Mira had disappeared in the event, but IR observations
again made it clear that the Mira was essentially unaffected by the white
dwarf outburst (Feast et al.\ 1977; 1983). However, there are some
peculiarities of the Mira. It is somewhat redder than you would expect for a
star with a 387 day period. It is a high latitude source at $b=-32^{\rm o}$,
so presumably the reddening is circumstellar. Interestingly, it also shows
what might be an ``eclipse'' in the $J$ light, around JD 2447500. However,
this event is followed by a lot of irregularity, another dip around JD
2449000 and possibly another one in progress at the moment. This is much
greater irregularity than was evident at earlier times. It is also
interesting to see that $L$ does not fade with $J$, but actually brightens a
little later. It is probably significant that we see these ``eclipses'' or
{\it obscuration events} in all symbiotic Mira that have been monitored for
more than about 10 yr; they have not been properly explained as yet
(Whitelock 1987). Their spectral signature certainly suggests dust, but is
it new dust ejected from the Mira, as we think is occuring for the carbon
stars, or is it old dust in orbit, as has been suggested for R Aqr? We
really need other types of monitoring over similar long time scales to sort
out what's going on in these stars.

\begin{figure}[t] 
\plotfiddle{fig6.ps}{7cm}{0}{90}{70}{-244}{-40}
\caption{The $K$ light curve for the carbon-rich Mira in the Sagittarius
Dwarf Galaxy. The observations cover 3.4 cycles of the 300 day period.} 
\end{figure}

\section{Miras in the Sagittarius Dwarf Galaxy} 
 I would like to finish by showing the light curve of a newly discovered
Mira in the Sagittarius dwarf galaxy. The Sagittarius dwarf was discovered
in 1994 by Ibata, Gilmore \& Irwin (1994).  It is at a distance of only 25
kpc from the sun being about 16 kpc behind the Galactic Centre. It appears
to be undergoing tidal disruption and to be in the process of merging with
the Milky Way. It contains planetary nebulae (Zijlstra \& Walsh 1997) and at
least one carbon Mira (Whitelock, Irwin \& Catchpole 1996). This is a
particularly useful variable as it provides us with the only carbon Mira
outside of the LMC with a well determined distance. Fig 6 illustrates its
$K$ light curve phased at the period of 300 day. The amplitudes of the
variations are $\Delta J=1.0$ mag, $\Delta H=0.7$ mag and $\Delta K=0.4$
mag; there can be no doubt that this is a large amplitude variable. The
distance modulus derived from these observations and the Groenewegen \&
Whitelock (1997) PL relation is $17.14\pm0.25$ mag (with essentially all the
uncertainty coming from the spread in the PL relation). This can be compared
with the distance modulus of $17.02\pm0.19$ mag which Mateo et al.\ (1995)
derive from the horizontal branch. There are several other luminous and red
carbon stars in the Sgr dwarf which we are monitoring from SAAO and which
may well be large amplitude variables.

\acknowledgments I am grateful to my colleagues from SAAO for allowing me to
use the IR data in advance of publication and to Michael Feast for many
fruitful discussions.


\begin{references}
\reference Carter, B.S. 1990, \mnras, 242, 1
\reference Caswell, J.L., Haynes, R.F. 1975, \mnras, 173, 649
\reference Feast, M.W. 1963, \mnras, 125, 367
\reference Feast, M.W. 1997, \mnras, 285, 339
\reference Feast, M.W., Robertson, B.S.C., Catchpole, R.M. 1977, \mnras, 
  179, 499
\reference Feast, M.W., et al. 1983, \mnras, 202, 951
\reference Feast, M.W., et al. 1989, \mnras, 241, 375
\reference Fleischer, A.J., Gauger, A., Sedlmayr, E. 1995, \aap, 297, 543
\reference Groenewegen, M.A.T., Whitelock, P.A. 1996, \mnras, 281, 1347
\reference H\"ofner, S., Dorfi, E.A. 1997, \aap, 319, 648
\reference H\"ofner, S., Feuchtinger, M.U., Dorfi, E.A. 1995, \aap, 297, 815
\reference Hollis, J.M., Pedelty, J.A.,  Lyon, R.G. 1997, \apj, L85
\reference Ibata, R.A., Gilmore, G., Irwin, M.J. 1994, Nature, 370, 194
\reference Iben, I., Renzini, A. 1983, \araa, 21, 271
\reference Kholopov, P.N. 1985, {\it General Catalogue of Variable Stars}, 
  4th edition, Nauka, Moscow (GCVS)
\reference Little, S.J., Little-Marenin, I.R., Bauer, W.H. 1987, \aj, 94,
  981
\reference Mateo, M., et al. 1995, \aj, 110, 1141
\reference Menzies, J.W., Whitelock, P.A. 1985, \mnras, 212, 783 
\reference Merrill, P.W. 1952, \apj, 116,21
\reference Renzini, A. 1981, in: (eds.) Iben I. \& Renzini A., {\it Physical
  Processes in Red Giants}, p.165, Dordrecht: Reidel
\reference van Leeuwen, F., Feast, M.W., Whitelock, P.A., Yudin, B. 1997,
  \mnras, 287 955
\reference Webster, B.L., Allen, D.A. 1975, \mnras, 171, 171
\reference Whitelock, P.A. 1987, \pasp, 99, 573
\reference Whitelock, P.A. 1996, in: (eds.) Sterken C. \& Jaschek C.,
{\it Light Curves of Variable Stars}, p.106, Cambridge: CUP
\reference Whitelock, P.A. 1997, in: (ed.) Wing R., {\it The Carbon Star
Phenomenon}, IAU Symp.\ 177, in press.
\reference Whitelock, P.A., Feast, M.W., Catchpole, R.M. 1991, \mnras, 248,
  276
\reference Whitelock, P.A., Irwin, M., Catchpole, R.M. 1996, New Astro., 1,
  57
\reference Whitelock, P.A., et al. 1983, \mnras, 202, 951
\reference Whitelock, P.A., et al. 1994, \mnras, 267, 711
\reference Whitelock, P.A., et al. 1997, \mnras, 288, 512
\reference Willson, L.A., Garnavich, P., Mattei, J.A. 1981, IBVS, No.\ 1961
\reference Wilson, W.J., Barrett, A.H. 1972, \apj, 177, 523
\reference Winters, J.M., et al. 1994, \aap, 290, 623 
\reference Zijlstra, A.A., Walsh, J.A. 1997, \aap, in press
\end{references}
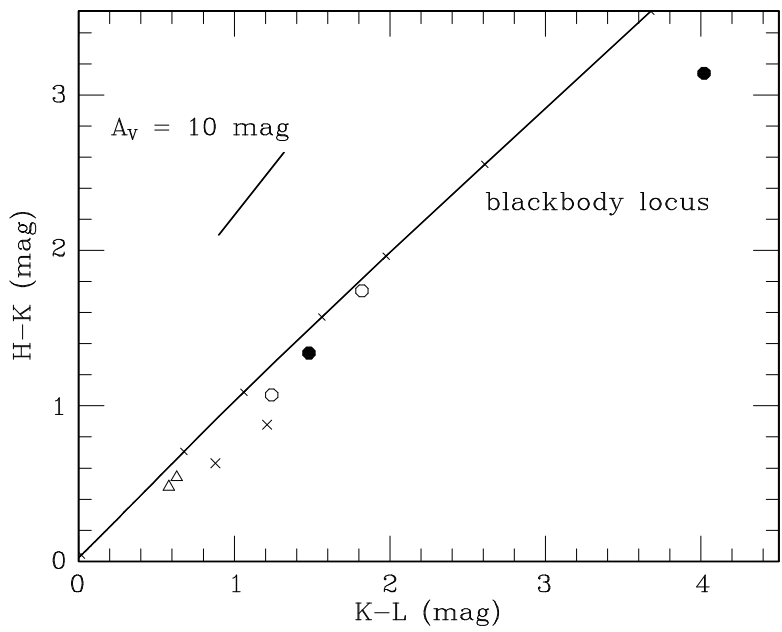
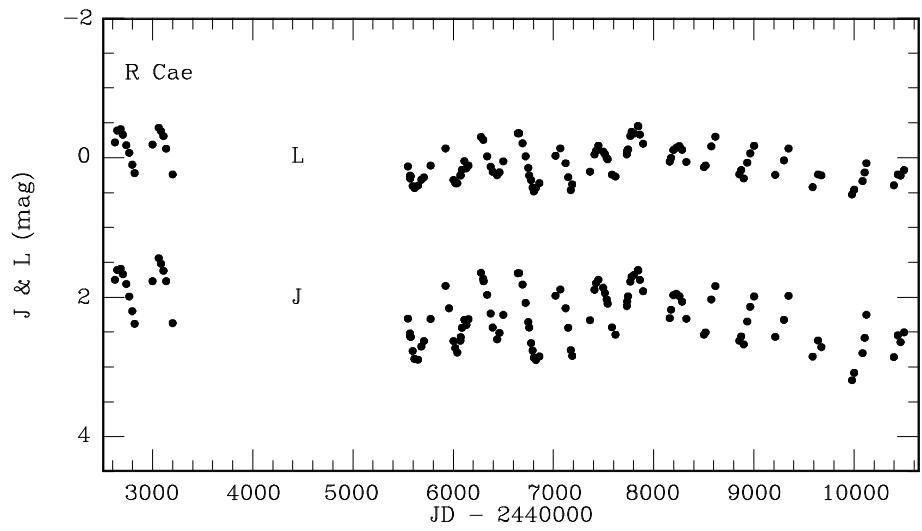
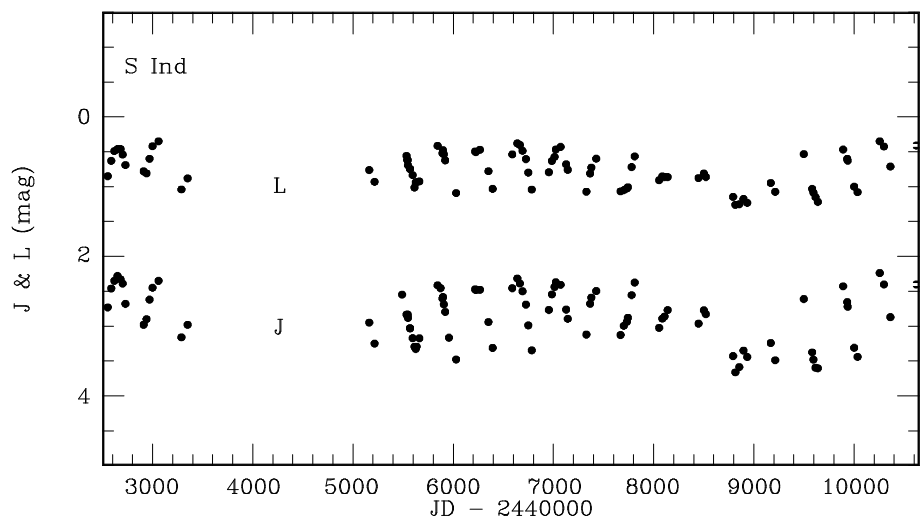
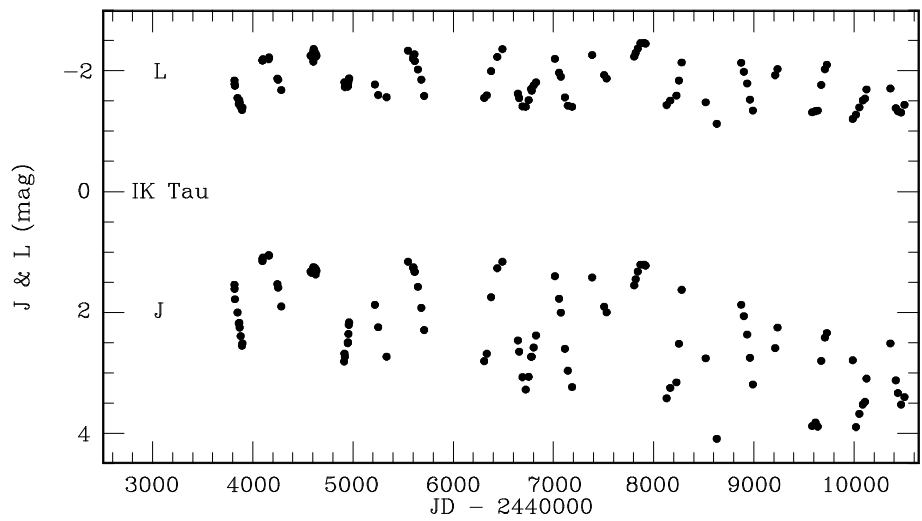
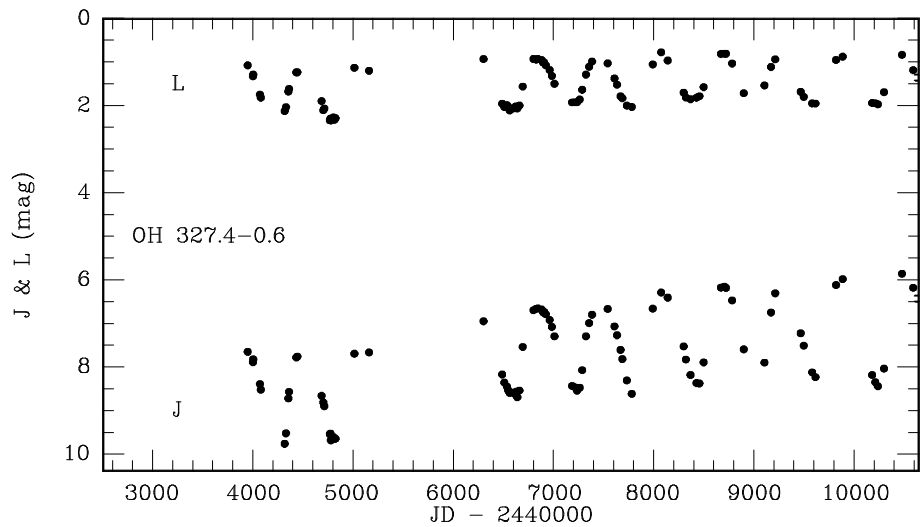
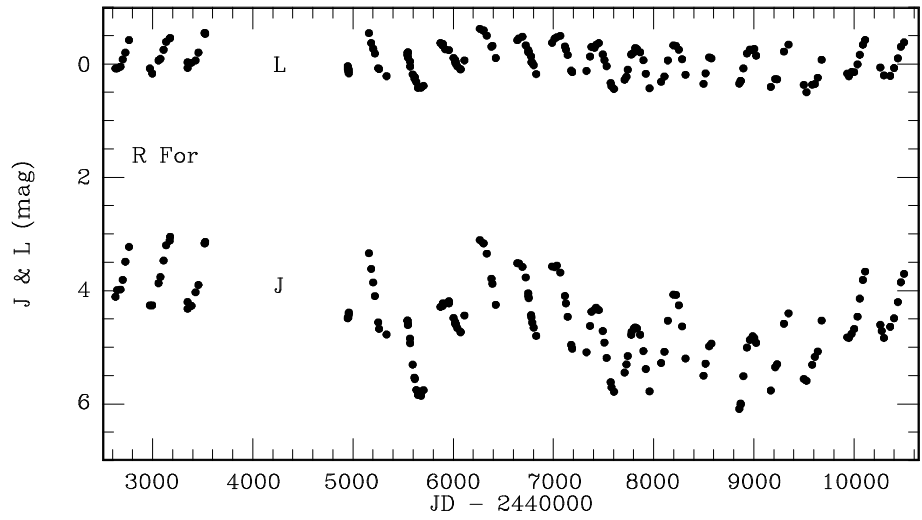
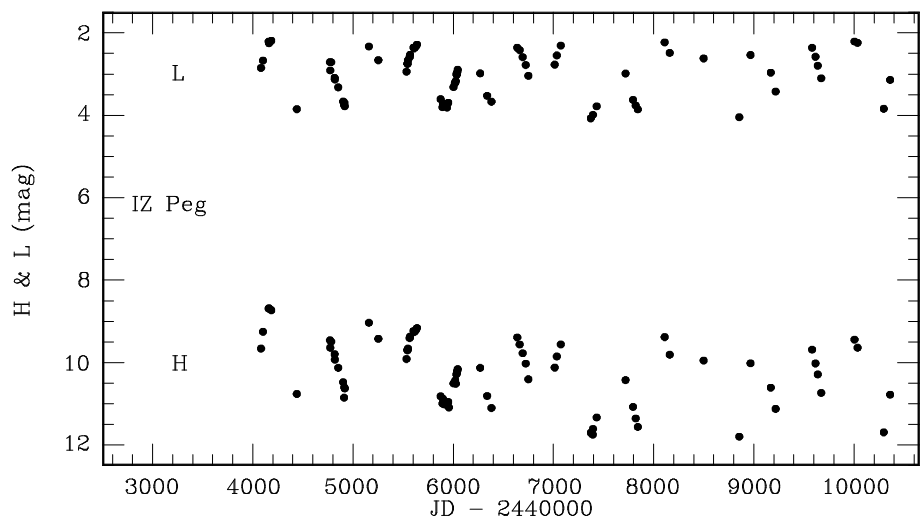
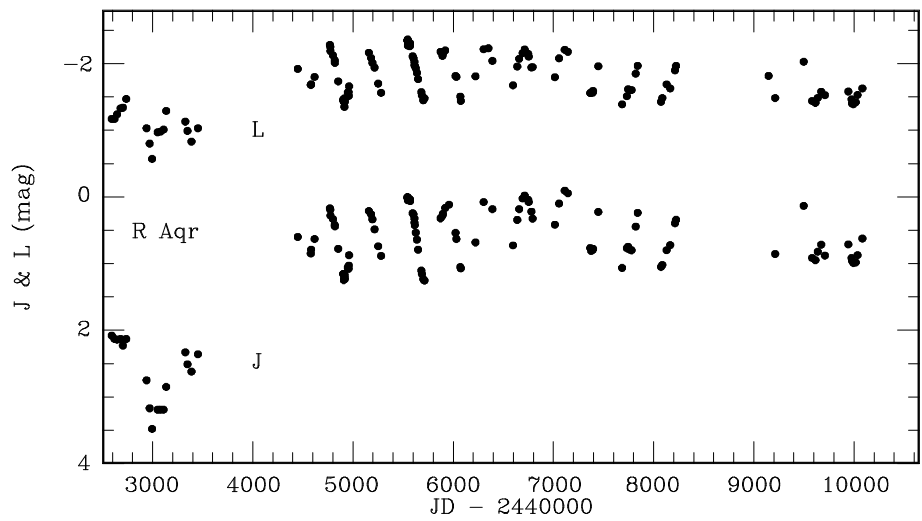
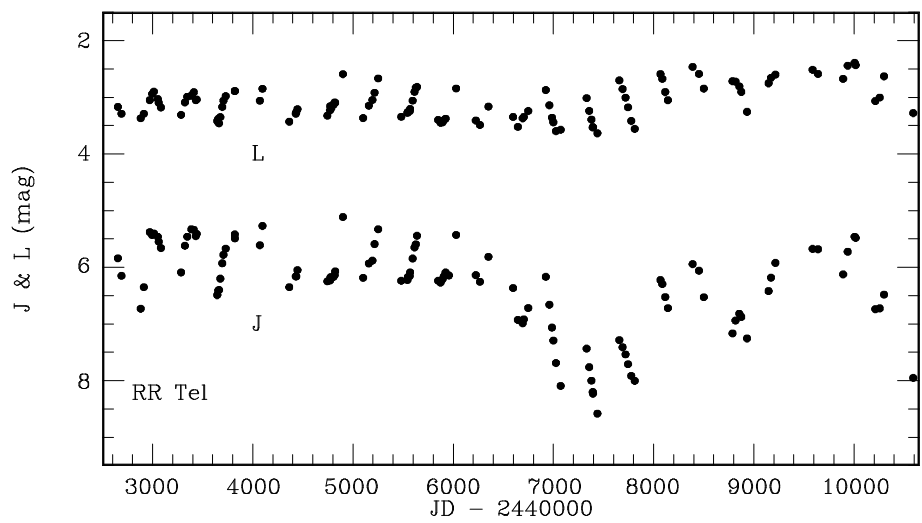
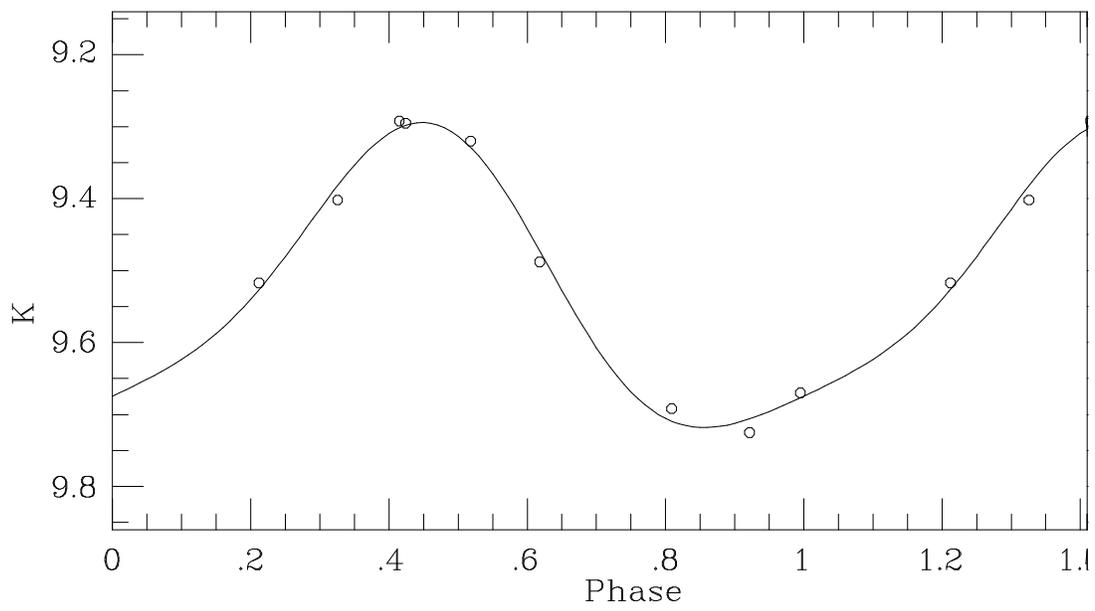
\end{document}